\documentclass[pre,epsfig,twocolumn,showpacs,preprintnumbers,amssymb]{revtex4}
\usepackage{bm,graphicx,amsmath}

\begin{document}
\title{Persistent random walk on a site-disordered one-dimensional lattice: Photon subdiffusion}
\author{MirFaez Miri}
\email{miri@iasbs.ac.ir}
\author{Zeinab Sadjadi}
\affiliation{Institute for Advanced Studies in Basic Sciences,
Zanjan, 45195-1159, Iran}
\author{M Ebrahim Fouladvand}
\affiliation{Department of Physics, Zanjan University, Zanjan, P.O. Box 313, Iran}

\begin{abstract}
We study the persistent random walk of photons on a one-dimensional lattice of random transmittances.
Transmittances at different sites are assumed independent, distributed according to a given probability density $f(t)$.
Depending on the behavior of $f(t)$ near $t=0$, diffusive and subdiffusive transports are predicted by the
disorder expansion of the mean square-displacement and the effective medium approximation. 
Monte Carlo simulations confirm the anomalous diffusion of photons. 
To observe photon subdiffusion experimentally, we suggest a dielectric film stack
for realization of a distribution $f(t)$.
\end{abstract}

\maketitle

\section{Introduction}\label{intro}
Random walks figure prominently in a multitude of different physical problems.
This is exemplified by such diverse fields
as the polymer physics\ \cite{doi}, crystallographic statistics\ \cite{weiss2,weiss},
transport in disordered media\ \cite{weiss,kehr,Bouchaud},
bacterial motion, and other types of biological migration\ \cite{berg}.

Random walks with {\it correlated} displacements, model a host of phenomena.
For example, the vacancy mechanism of atom diffusion in solids incorporates a
correlation effect, since an atom has a larger probability to move backward to the hole it just vacated
rather than onward\ \cite{bardeen}. Correlations also arise in diffusion of guest molecules in
zeolite channels\ \cite{zeo}, electron hopping in Coulomb glass\ \cite{glass}, motion of excitons
at low temperatures in mixed naphthalene crystals\ \cite{nap}, etc.
Among the correlated walks, the {persistent random walk} is possibly
the simplest one to incorporate a form of momentum in addition to random
motion\ \cite{weiss,kehr}.
In its basic realization on a one-dimensional
lattice, a persistent random walker possesses constant probabilities
for either taking a step in the same direction as the immediately
preceding one or for reversing its motion.
Firstly introduced by F\"{u}rth as a model for diffusion in a number of
biological problems\ \cite{furth}, and shortly after by Taylor in the
analysis of turbulent diffusion\ \cite{tay}, the persistent random walk
model is now generalized to study, e.g., polymers\ \cite{godoy}, chemotaxis\ \cite{chemotaxis}, cell movement\ \cite{cell},
and general transport mechanisms\ \cite{t0,www}.

Recently, the persistent random walk model is utilized in the
description of diffusive light transport in foams\ \cite{miri1,miri2,miri3,miri4,miri5} which is well established by experiments\ \cite{dur}.
A relatively dry foam consists of cells separated by thin liquid films\ \cite{Weaire1999}.
Cells in a foam are much larger than the wavelength of light, thus one can employ ray optics and
follow a light beam or photon as it is transmitted through the liquid films
with a probability $t$ called the intensity transmittance. This
naturally leads to a persistent random walk of the photons.
Special attention is paid to the light transport in the ordered honeycomb and Kelvin structures, which have
been used for an analytic access to the physical properties of disordered foams
as exemplified by work on their rheological behavior\ \cite{1982}.
Symmetries of hexagonal and tetrakaidecahedral cells allow identifying specific
one- and two-dimensional random walks, which are absent in disordered foams\ \cite{miri1, miri4}.
Moreover, in a first model, it is assumed that intensity transmittance does not depend on the
incidence angle of photons and film thickness. Nevertheless, analytical treatment of these peculiar and simple
random walks facilitates interpretation of
Monte Carlo simulations, in which topological disorder of foams and exact thin-film intensity transmittance
are taken into account\ \cite{miri2,miri4}.

In the ordered honeycomb (Kelvin) structure,
the one-dimensional persistent walk arises when the photons move perpendicular to a cell edge (face).
Thin-film transmittance depends on the film thickness. Films are not expected to have the same thickness.
These observations motivates us to consider persistent random walk on a one-dimensional lattice of {\it random} transmittances.
We assume that transmittances at different sites are independent random variables, distributed according to a given probability density $f(t)$.

Our first approach to the problem is based upon a disorder expansion of the mean square-displacement due to
Kundu, Parris, and Phillips\ \cite{K}. Assuming that $<{1}/{t}>=\int_{0}^{1} {f(t)}/{t} ~dt$
is finite,
we validate the classical persistent random walk with an effective transmittance $t_{eff}$, where
$1/t_{eff}=<1/t>$.
Inspired by this result, we generalize the effective medium approximation (EMA)
formulated by Sahimi, Hughes, Scriven, and Davis\ \cite{sahimi, emabooks}
to investigate the transport on a line with infinite $<1/t>$.
We show that if $f(t) \rightarrow f(0) $ as $t \rightarrow 0 $, the mean square-displacement after
$n$ steps is proportional to $n/\ln(n)$. If $f(t) \sim f_{\alpha} t^{-\alpha} $ ($0<\alpha<1$) as
$t \rightarrow 0 $, we find that the mean square-displacement is proportional to $ n^{(2-2\alpha)/(2-\alpha)}  $. 
Our Monte Carlo simulations confirm
the {anomalous diffusion of photons}. Quite interesting, we find that anomalous diffusion of persistent walkers
and {\it hopping} particles on a site-disordered lattice\ \cite{sahimi, emabooks, alex} are similar.
Finally, for the experimental observation of the photon subdiffusion, we suggest a dielectric film stack
to realize small transmittances.

Our paper is organized as follows. In Sec.\ \ref{model} we introduce the model.
The perturbative approach and the effective medium approximation are discussed in
Secs.\ \ref{mypertu} and\ \ref{myema}, respectively. The numerical treatment and its
results are reported in Sec.\ \ref{montecarlo}. We close with a discussion of our results and conclusions in
Sec.\ \ref{discuss}.

\section{Model}\label{model}

We consider a one-dimensional lattice random walk in which steps are permitted to the nearest
neigbor sites only. We normalize the length and duration of a step to 1.
At each site $\bm{j}$, a walker either takes a step in the same direction as
the immediately preceding one with a probability $t_{\bm{j}}$, or reverses its motion with a
probability $r_{\bm{j}}=1- t_{\bm{j}}$. Here we assume symmetric transmittances, i.e.
$t_{\bm{j} \rightarrow \bm{j}+1 }= t_{\bm{j} \rightarrow \bm{j}-1 } = t_{\bm{j}}$, as
the transmittance of a thin-film is the same whether the light ray is going to the right ($+$)
or to the left ($-$) direction.

We assume that i) transmittance at each site is a {random} variable, ii) transmittances at two
different sites are independent, iii) transmittances at all sites are
distributed according to a given normalized probability density $f(t)$. Apparently
$\int_{0}^{1}   f(t) dt=1 $. For any function $h(t)$, we define  $<h(t)>=\int_{0}^{1}  h(t) f(t) dt$.

We denote by $P^{+}(n, \bm{j})$ $\big ( P^{-}(n, \bm{j}) \big)$ the probability
that the walker after its $n$th step arrives at site $\bm{j}$ with positive (negative) momentum.
A set of two master equations can be established to couple the probabilities at step $n+1$
to the probabilities at step $n$:
\begin{eqnarray}
P^{+}(n+1, \bm{j})&=& t_{\bm{j}-1}  P^{+}(n, \bm{j}-1) + r_{\bm{j}-1}  P^{-}(n, \bm{j}-1) ,\nonumber \\
P^{-}(n+1, \bm{j})&=&  r_{\bm{j}+1} P^{+}(n, \bm{j}+1)  +t_{\bm{j}+1} P^{-}(n, \bm{j}+1) . \label{master1} \nonumber \\
\end{eqnarray}

For the description of the photon distribution on the line, we do not
need to specify the internal state ($\pm$) explicitly.
That means we are mainly interested in the probability that the photon
arrives at position $\bm{j}$ at step $n$,
\begin{equation}
P(n, \bm{j})=P^{+}(n, \bm{j})+P^{-}(n, \bm{j}), \label{sum1}
\end{equation}
from which we extract the first and second moments after $n$ steps
as the characteristic features of a random walk:
\begin{eqnarray}
 \langle \langle {\bm{j} }\rangle   \rangle _n  &=& \langle\sum_{\bm{j}}        \bm{j}    P(n, \bm{j})  \rangle ,\nonumber \\
\langle  \langle {\bm{j}^2 }\rangle \rangle _{n}   &=&\langle \sum_{\bm{j}}   \bm{j}^2 P(n, \bm{j}) \rangle .
\label{moments}
\end{eqnarray}
Here the first bracket represents an ensemble average over all random transmittances,
and the second bracket signifies an average with respect to the distribution $P(n, \bm{j})$.

One obtains the classical persistent random walk assuming a constant transmittance $t$ at each site.
Translational invariance of the medium is then invoked to deduce the exact solution of $P(n, \bm{j})$
in the framework of characteristic functions
(the spatial Fourier transforms of probability distributions)\ \cite{weiss,kehr}. Furthermore,
the mean square-displacement of photons after
$n \rightarrow \infty $ steps can be obtained as
\begin{equation}
\langle {\bm{j}^2 }\rangle_n= \frac{t}{1-t} n. \label{dif1}
\end{equation}

Considering a lattice of random transmittances with an almost narrow distribution $f(t)$,
one may intuitively expect normal diffusion of the photons, and the validity of Eq. (\ref{dif1}) with an
effective transmittance $<t>$.
However, a closer inspection reveals that even a few sites with small transmittances (large reflectances) may
drastically hinder the photon diffusion: In the extreme limit where two transmittances are zero,
photons are confined between two sites.
This peculiar aspect of diffusion on a {\it one}-dimensional lattice, rules out the above guess.
In the following section, we present a sound perturbative approach to the problem.

\section{ Systematic disorder expansion of mean square-displacement of the photons}\label{mypertu}

Many of the approaches to the transport in disordered media have the disadvantage of being
restricted to one-dimensional problems.
Here we adopt the method of Kundu et al. \ \cite{K}, which is
applicable to two- and three-dimensional media.
We obtain a series of approximate solutions for photon transport on a disordered line
by transforming the master equation (\ref{master1}) to an equivalent but
more appropriate integral equation for the characteristic function
\begin{equation}
P^{\pm}(n, \bm{\theta})=\sum_{\bm{j} =-\infty}^{\infty}  P^{\pm}(n, \bm{j}) e^{i \bm{j} \bm{\theta}}. \label{chac1}
\end{equation}
This crucial step can be achieved utilizing the generalized generating functions
\begin{equation}
{\widetilde{\mathbf{P}}}^{\pm}(n, \bm{\theta})=\sum_{\bm{j} }  t_{\bm{j}}  P^{\pm}(n, \bm{j}) e^{i \bm{j} \bm{\theta}},
\label{chac2}
\end{equation}
as will be shown in the following.

First we simplify the set of coupled linear difference equations (\ref{master1}) using the method
of the $z$-transform\ \cite{weiss, jury} explained in Appendix\ \ref{zT}:
\begin{eqnarray}
\frac{P^{+}(z, \bm{j})}{z} -   \frac{P^{+}(n=0, \bm{j})}{z}&=&
t_{\bm{j}-1}  P^{+}(z, \bm{j}-1)\nonumber \\
& & + r_{\bm{j}-1}  P^{-}(z, \bm{j}-1) ,\nonumber \\
 \frac{P^{-}(z, \bm{j})}{z}   - \frac{P^{-}(n=0, \bm{j})}{z}&=&
 r_{\bm{j}+1} P^{+}(z, \bm{j}+1) \nonumber \\
& & +t_{\bm{j}+1} P^{-}(z, \bm{j}+1) . \label{master2}
\end{eqnarray}
We assume the initial conditions $P^{+}(n=0, \bm{j})=P^{-}(n=0, \bm{j})=\delta_{\bm{j},0}/2$.
Now we multiply both sides of (\ref{master2}) with $\exp({i \bm{j} \bm{\theta}} )$ and sum over $j$, which leads to
\begin{equation}
\mathbf{M}_3 (z,\bm{\theta} )\begin{pmatrix} P^{+}(z,\bm{\theta})\\P^{-}(z,\bm{\theta})\end{pmatrix} =
\mathbf{M}_2(\bm{\theta} )\begin{pmatrix} \widetilde{\mathbf{P}}^{+}(z, \bm{\theta})\\\widetilde{\mathbf{P}}^{-}(z, \bm{\theta})\end{pmatrix}
+\mathbf{M}_1(z ) \\  \label{mastermatrix},
\end{equation}
where
\begin{eqnarray}
\mathbf{M}_3 (z,\bm{\theta} )&=& \begin{pmatrix} \frac{1}{z} & -e^{i \bm{\theta}} \\ -e^{-i \bm{\theta}}& \frac{1}{z} \end{pmatrix} ,\nonumber \\
\mathbf{M}_2 (\bm{\theta} )&=&\begin{pmatrix} e^{i \bm{\theta}}& -e^{i \bm{\theta}}\\-e^{-i \bm{\theta}}&e^{-i \bm{\theta}}\end{pmatrix} ,\nonumber \\
\mathbf{M}_1 (z)&=&\begin{pmatrix}\frac{1}{2z}\\\frac{1}{2z}\end{pmatrix}.
\end{eqnarray}
Thus $P^{\pm}(z,\bm{\theta} )$ can be easily expressed
in terms of $\widetilde{\mathbf{P}}^{\pm}(z, \bm{\theta})$. 
From Eqs. (\ref{chac1}) and (\ref{chac2})
\begin{eqnarray}
 P^{\pm}(z, \bm{j})=\frac{1}{2 \pi t_{\bm{j}}} \int_{-\pi}^{\pi} {\widetilde{\mathbf{P}}}^{\pm}(z, \bm{\phi})
e^{-i \bm{j} \bm{\phi }}  d\bm{\phi} ,& & \nonumber \\
 P^{\pm}(z, \bm{\theta})=\sum_{\bm{j} } \frac{1}{2 \pi t_{\bm{j}}} \int_{-\pi}^{\pi}
{\widetilde{\mathbf{P}}}^{\pm}(z, \bm{\phi})
e^{-i \bm{j} (\bm{\phi }- \bm{\theta}) }  d\bm{\phi}, & & \label{chac3}
\end{eqnarray}
hence we can immediately transform Eq. (\ref{mastermatrix}) to an equivalent integral equation for
$\widetilde{\mathbf{P}}^{\pm}(z, \bm{\theta})$. It reads
\begin{widetext}
\begin{equation}
(<\frac{1}{t}> \mathbf{I} - \mathbf{M}_3^{-1} (z,\bm{\theta} )\mathbf{M}_2(\bm{\theta} ) )
\begin{pmatrix} \widetilde{\mathbf{P}}^{+}(z, \bm{\theta})\\\widetilde{\mathbf{P}}^{-}(z, \bm{\theta})\end{pmatrix}
=\mathbf{M}_3^{-1}(z,\bm{\theta} )\mathbf{M}_1(z)
- \sum_{\bm{j}} \frac{\Delta_{\bm{j} }  }{2 \pi} \int_{-\pi}^{\pi} e^{-i \bm{j} (\bm{\phi }- \bm{\theta}) }
\begin{pmatrix} \widetilde{\mathbf{P}}^{+}(z, \bm{\phi})\\\widetilde{\mathbf{P}}^{-}(z, \bm{\phi})\end{pmatrix} d\bm{\phi} ,
\label{masterintegral}
\end{equation}
\end{widetext}
where the random variable $ \Delta_{\bm{j} }$ is defined as
\begin{equation}
\Delta_{\bm{j}}= \frac{1}{t_{\bm{j}}  }-<\frac{1}{t}>, \label{deldef}
\end{equation}
and $\mathbf{I}$ is the identity matrix.
Note that two assumptions, that none of the transmittances is zero, and $<{1}/{t}>$ is finite, are required for
the validity of Eqs. (\ref{chac3})-(\ref{deldef}).

Successive approximations to the solution of integral equation (\ref{masterintegral}) can be generated by the iteration method.
Then ensemble averaging over the random transmittances is straightforward.
We use the identity $<\Delta_{\bm{j}} >=0$ and the assumption
\begin{equation}
< \Delta_{\bm{j}}\Delta_{\bm{j'}}>= \delta_{\bm{j},\bm{j'} }  \Delta^2 ,
\label{Delta2}
\end{equation}
to obtain the leading terms:
\begin{equation}
<\begin{pmatrix} \widetilde{\mathbf{P}}^{+}(z, \bm{\theta})\\\widetilde{\mathbf{P}}^{-}(z, \bm{\theta})\end{pmatrix} >
=\big( 1+  \Delta^2 \mathbf{M}_4(z,\bm{\theta} )  \big )
\begin{pmatrix} \widetilde{\mathbf{P}}_{0}(z, \bm{\theta})\\\widetilde{\mathbf{P}}^{*}_{0}(z, \bm{\theta})\end{pmatrix},
\label{solution}
\end{equation}
where matrix $ \mathbf{M}_4(z,\bm{\theta} )$ is given in
Appendix\ \ref{apMatrix},
\begin{equation}
\widetilde{\mathbf{P}}_{0}(z, \bm{\theta})=
\frac{1}{2 <\frac{1}{t}>}
\frac{ z( <\frac{1}{t}> e^{i \bm{\theta}} -2 \cos\bm{\theta}) +<\frac{1}{t}>   }{z^2(2-<\frac{1}{t}> )  -2z\cos\bm{\theta}  +  <\frac{1}{t}>},
\end{equation}
and $*$ denotes the complex conjugation.

The task is now calculating $< P^{\pm}(z,\bm{\theta} )>$ from Eq. (\ref{mastermatrix}),
and using the identity
\begin{equation}
\sum_{\bm{j}}   \bm{j}^l  P^{\pm}(z, \bm{j})=\frac{1}{i^l}
\frac{\partial^l P^{\pm}(z, \bm{\theta})}{\partial\bm{\theta}^l } |_{ \bm{\theta}=0}
~~~~( l= 1,2,...),\label{help1}
\end{equation}
to obtain mean square-displacements of the photons.
To this end,
\begin{eqnarray}
 & \sum \langle  \langle {\bm{j}^2 }\rangle \rangle _{n}  z^n & \nonumber \\
&= &\frac{z-4iz \big (<\frac{\partial \widetilde{\mathbf{P}}^{+}(z, \bm{\theta}) }{\partial\bm{\theta} }-\frac{\partial\widetilde{\mathbf{P}}^{-}(z, \bm{\theta}) }{\partial\bm{\theta} }  >\big)|_{\bm{\theta}=0} }{1-z^2}, \nonumber \\
             & \sim&  \frac{1}{<\frac{1}{t}>-1}  \frac{1}{(1-z)^2}  \nonumber \\
 & +&  \frac{\sqrt{2} \Delta^2}{2  (<\frac{1}{t}>-1)^{2.5}}\frac{1}{(1-z)^{1.5}},
\end{eqnarray}
as $ z \rightarrow 1 $. We use Tauberian theorem (\ref{Tau1}) to conclude
\begin{equation}
\langle  \langle {\bm{j}^2 }\rangle \rangle _{n}
 =
 \frac{1}{<\frac{1}{t}>-1} n+
\frac{\sqrt{2} \Delta^2}{\sqrt{\pi}  (<\frac{1}{t}>-1)^{2.5}} \sqrt{n}. \label{dfg}
\end{equation}
Mean square-displacement of the photons after $n \rightarrow \infty $ steps is indeed proportional to $n$.
We validate the classical persistent random walk result (\ref{dif1}) with
an effective transmittance $t_{eff}$, where
\begin{equation}
\frac{1}{t_{eff}}=<\frac{1}{t}>. \label{tefff}
\end{equation}

\section{Effective Medium Approximation}\label{myema}
Stochastic transport in random media is often subdiffusive or
superdiffusive\ \cite{weiss,kehr,Bouchaud, emabooks}.
At this stage, we pay attention to persistent random walk on a line with infinite $<1/t>$,
where the normal diffusion is not guaranteed. Our approach to the problem is based upon
a variant of effective medium approximation (EMA) developed in\ \cite{sahimi, sahimi2}.

To facilitate solution of Eq. (\ref{master2}) we introduce a
reference lattice or average medium, with all intensity transmittances (reflectances) equal to
$t_{e} (z) $ $\big ( r_{e} (z) \big )$, and probabilities $P^{\pm}_{e}(z, \bm{j})$, so that
\begin{eqnarray}
\frac{P^{+}_{e}(z, \bm{j})}{z} -   \frac{P^{+}(n=0, \bm{j})}{z}&=&
t_{e} (z)  P^{+}_{e}(z, \bm{j}-1)\nonumber \\
& & + r_{e} (z)   P^{-}_{e}(z, \bm{j}-1) ,\nonumber \\
 \frac{P^{-}_{e}(z, \bm{j})}{z}   - \frac{P^{-}(n=0, \bm{j})}{z}&=&
 r_{e} (z) P^{+}_{e}(z, \bm{j}+1) \nonumber \\
& & +t_{e} (z)P^{-}_{e}(z, \bm{j}+1) . \label{masterema}
\end{eqnarray}
EMA determines $t_{e} (z) $ and $ r_{e} (z) $ in a self-consistent manner, in which the role of
distribution $f(t)$ is manifest. This is done by taking a cluster of random transmittances from the
original distribution, and embedding it into the effective medium.
We then require that {\it average} of site occupation probabilities of the decorated medium duplicate
$P^{\pm}_{e}(z, \bm{j})$ of the effective medium. We will sketch the method in the following.

Subtracting Eqs. (\ref{master2}) and (\ref{masterema}), we obtain
\begin{widetext}
\begin{eqnarray}
& &
\frac{1}{z}\begin{pmatrix} Q^{+}(z,\bm{j})\\Q^{-}(z,\bm{j})\end{pmatrix}
-\mathbf{T}^{-}(z)\begin{pmatrix} Q^{+}(z,\bm{j-1})\\Q^{-}(z,\bm{j-1})\end{pmatrix}
-\mathbf{T}^{+}(z)\begin{pmatrix} Q^{+}(z,\bm{j+1})\\Q^{-}(z,\bm{j+1})\end{pmatrix}= \nonumber \\
& & \big [  \begin{pmatrix} t_{\bm{j-1}}  & r_{\bm{j-1}} \\ 0 & 0 \end{pmatrix} -\mathbf{T}^{-}(z)\big]
\begin{pmatrix} P^{+}(z,\bm{j-1})\\P^{-}(z,\bm{j-1})\end{pmatrix}+
\big[  \begin{pmatrix}  0 & 0 \\ r_{\bm{j+1}} & t_{\bm{j+1}} \end{pmatrix}-\mathbf{T}^{+}(z) \big]
\begin{pmatrix} P^{+}(z,\bm{j+1})\\P^{-}(z,\bm{j+1})\end{pmatrix},
\label{mainema}
\end{eqnarray}
\end{widetext}
where
\begin{eqnarray}
\begin{pmatrix} Q^{+}(z,\bm{j})\\Q^{-}(z,\bm{j})\end{pmatrix} & =&
\begin{pmatrix} P^{+}(z,\bm{j})\\P^{-}(z,\bm{j})\end{pmatrix}
-\begin{pmatrix} P^{+}_{e}(z,\bm{j})\\P^{-}_{e}(z,\bm{j})\end{pmatrix},\nonumber \\
 \mathbf{T}^{-}(z) & =& \begin{pmatrix}  t_{e}(z) &r_{e}(z) \\ 0 & 0 \end{pmatrix},\nonumber \\
 \mathbf{T}^{+}(z) & =& \begin{pmatrix}  0 & 0 \\ r_{e}(z) & t_{e}(z) \end{pmatrix}.
\end{eqnarray}
Equation (\ref{mainema}) suggests to define an associated Green function
$\mathbf{G}(z,\bm{j})= \begin{pmatrix} G_{11} & G_{12} \\ G_{21} & G_{22}  \end{pmatrix}$
by the equation
\begin{eqnarray}
& &\frac{1}{z} \mathbf{G}(z,\bm{j})
-\mathbf{T}^{-}(z)   \mathbf{G}(z,\bm{j}-1)
-\mathbf{T}^{+}(z)    \mathbf{G}(z,\bm{j}+1) = \delta_{\bm{j},0} \mathbf{I} ,\nonumber \\
\end{eqnarray}
whose solution is the inverse Fourier transform of
\begin{eqnarray}
 \mathbf{G}(z,\bm{\theta})&=&\frac{z^2}{1-2z {t_{e}(z)} \cos\bm{\theta} + z^2[t^{2}_{e}(z) - r^{2}_{e}(z)]}\nonumber \\
& & \times  \begin{pmatrix} \frac{1}{z}- {t_{e}(z)} e^{-i\bm{\theta}} &  r_{e}(z) e^{i\bm{\theta}} \\
 r_{e}(z) e^{-i\bm{\theta}} &  \frac{1}{z}- {t_{e}(z)} e^{i\bm{\theta}}\end{pmatrix}.
\end{eqnarray}
For the present, we consider only the simplest approximation, and embed {\it one}
random transmittance at site $\bm{l}$ of the effective medium. Then solution of Eq. (\ref{mainema}) is
\begin{eqnarray}
\begin{pmatrix} Q^{+}(z,\bm{j})\\Q^{-}(z,\bm{j})\end{pmatrix} & = &
\int_{0}^{2 \pi} \mathbf{G}(z,\bm{\theta}) \mathbf{S}(z,\bm{\theta})  e^{-i\bm{\theta} (\bm{j} -\bm{l}) }\nonumber \\
& &~~~~~~\times \begin{pmatrix} P^{+}(z,\bm{l})\\P^{-}(z,\bm{l})\end{pmatrix} \frac{d \bm{\theta}}{ 2 \pi},
\label{poi}
\end{eqnarray}
where
\begin{eqnarray}
  \mathbf{S}(z,\bm{\theta})&=&
 \begin{pmatrix} [t_{\bm{l}} -t_{e}(z)] e^{i\bm{\theta}} &  [r_{\bm{l}} -r_{e}(z)] e^{i\bm{\theta}} \\
    [r_{\bm{l}} -r_{e}(z)] e^{-i\bm{\theta}}  & [t_{\bm{l}} -t_{e}(z)] e^{-i\bm{\theta}}\end{pmatrix}.
 \end{eqnarray}

Self-consistency equation is $ <P^{\pm}(z,\bm{l}) >=P^{\pm}_{e}(z,\bm{l})$, or
\begin{equation}
< \big [\mathbf{I}-\int_{0}^{2 \pi} \mathbf{G}(z,\bm{\theta}) \mathbf{S}(z,\bm{\theta})
\frac{d \bm{\theta}}{ 2 \pi} \big]^{-1} >=\mathbf{I}.
\end{equation}
The above matrix equation leads to two independent self-consistency conditions.
Choosing $t_{e}(z)+r_{e}(z)=1$, one of the conditions can be satisfied. It is a signature of
the success of effective medium to represent the original random medium,
where $t_{\bm{j}}+ r_{\bm{j}}=1$ holds at any site $\bm{j}$.
The second self-consistency condition then determines $t_{e}(z)$:
\begin{equation}
\frac{1}{t_{e}(z)}=\int_{0}^{1} \frac{f(t)  dt}{[t_{e}(z)-t] \frac{\sqrt{1-z^2}}{ \sqrt{1-[2t_{e}(z)-1]^2 z^2 }  } +t} ,
\label{selfconsis}
\end{equation}
in which the role of distribution $f(t)$ is manifest.

$Z$-transform of mean square-displacement of the photons in the effective medium
can be obtained as
\begin{equation}
\sum_{n=0}^{\infty}  \langle  \langle {\bm{j}^2 }\rangle \rangle _{n}  z^n=
\frac{z}{(1-z)^2} \frac{1+z [2t_{e}(z)-1]}{1-z[2t_{e}(z)-1]}.
\label{jema}
\end{equation}
We are interested in the long time behavior, thus Tauberian theorems suggest to analyze
Eqs. (\ref{selfconsis}) and (\ref{jema}) in the limit $ z \rightarrow 1$.

First we assume that $<{1}/{t}>$ is finite and $t_{e}(z)$ has no singularity
in the limit $ z \rightarrow 1$. Then Eq. (\ref{selfconsis}) yields
\begin{equation}
\frac{1}{t_{e}(z)}=\int_{0}^{1} \frac{f(t)  dt}{t} ,
\label{selfconsis2}
\end{equation}
in accordance with our second assumption. We deduce from
Eqs. (\ref{jema}) and (\ref{Tau1}) that
\begin{equation}
 \langle  \langle {\bm{j}^2 }\rangle \rangle _{n}
 = \frac{1}{<\frac{1}{t}>-1} n, \label{difema1}
\end{equation}
therefore the system evolves diffusively.
This is in complete agreement with the predictions of Sec.\ \ref{mypertu}.

When $<{1}/{t}>$ is infinite, the behavior of ${1}/{t_{e}(z)}$ is determined by the behavior of
$f(t)$ at small values of $t$. Let us assume that $f(t)$ has a finite derivative at $t=0$.
We can decompose the integral in Eq. (\ref{selfconsis}) into a sum:
\begin{eqnarray}
\frac{1}{t_{e}(z)} &=&
\int_{0}^{\epsilon} \frac{f(0)  dt}{[t_{e}(z)-t] \frac{\sqrt{1-z^2}}{ \sqrt{1-[2t_{e}(z)-1]^2 z^2 }  } +t} \nonumber \\
& &+\int_{0}^{\epsilon} \frac{[f(t)-f(0) ] dt}{[t_{e}(z)-t] \frac{\sqrt{1-z^2}}{ \sqrt{1-[2t_{e}(z)-1]^2 z^2 }  } +t} \nonumber \\
& &+\int_{\epsilon}^{1} \frac{f(t)  dt}{[t_{e}(z)-t] \frac{\sqrt{1-z^2}}{ \sqrt{1-[2t_{e}(z)-1]^2 z^2 }  } +t},
\label{decomposition}
\end{eqnarray}
where $\epsilon$ is a small number. By assumption $f(t)-f(0)= t f'(0)$ near $t=0$,
and the factor $t$ cancels the potential
singularity in the second term as $z \rightarrow 1 $. Indeed the only singular behavior
can come from the first term. Equation (\ref{decomposition}) then leads to
\begin{equation}
\frac{1}{t_{e}(z)} \approx -f(0) \ln \big[  \frac{t_{e}(z) \sqrt{1-z^2} }{\sqrt{1-[2t_{e}(z)-1]^2 z^2 } }  \big].
\end{equation}
Methods for the asymptotic solution of transcendental equations\ \cite{trans} are invoked to obtain
\begin{equation}
t_{e}(z)\approx \frac{-2 }{f(0) \ln(1-z)}.
\end{equation}
We deduce from Eqs. (\ref{jema}) and (\ref{Tau2}) that
\begin{equation}
\langle  \langle {\bm{j}^2 }\rangle \rangle _{n}
= \frac{2}{f(0)}  \frac{n}{\ln(n)}, \label{difema2}
\end{equation}
thus the transport is {subdiffusive}.

We now investigate the cases in which $f'(0)$ is infinite.
If $f(t) \sim f_{\alpha} t^{-\alpha} $ ($0<\alpha<1$) as
$t \rightarrow 0 $, self-consistency equation (\ref{selfconsis}) yields
\begin{equation}
[t_{e}(z)]^{\alpha-1}= \frac{\pi f_{\alpha} }{\sin(\pi \alpha )}
 \big[ \frac{1-[2t_{e}(z)-1]^2 z^2}{1-z^2}  \big]^{\frac{\alpha}{2}},
\end{equation}
where we have used $\int_{0}^{\infty} x^{-\alpha}/(1+x) dx= \pi /\sin(\pi \alpha)  $.
In this case we find
\begin{equation}
t_{e}(z) \approx
[ 2^{{-\alpha}/{2} }  \frac{\sin(\pi \alpha )}{\pi f_{\alpha} }  ]^{ \frac{2}{2-\alpha} }
~(1-z)^{  \frac{ \alpha}{2-\alpha}  },
\end{equation}
and
\begin{equation}
 \langle  \langle {\bm{j}^2 }\rangle \rangle _{n}
=\frac{ [ 2^{{-\alpha}/{2} }  \frac{\sin(\pi \alpha )}{ \pi f_{\alpha}}  ]^{ \frac{2}{2-\alpha} } }{\Gamma(\frac{4-3\alpha }{2-\alpha})}
~ n^{  \frac{2-2\alpha }{2-\alpha}}. \label{difema3}
\end{equation}
Apparently, $0<(2-2\alpha) /(2-\alpha)<1$ and the transport is {subdiffusive}.

\section{Numerical simulations}\label{montecarlo}

The predictions of EMA can be inspected by numerical simulations.
The computer program produces 400 media, whose transmittances are distributed according to a given
$f(t)$. For each medium, it takes $10^3$ photons at the initial position $\bm{j}=0$ and generates the trajectory
of each photon following a standard Monte Carlo procedure.

\begin{figure}
\includegraphics[width=0.85\columnwidth]{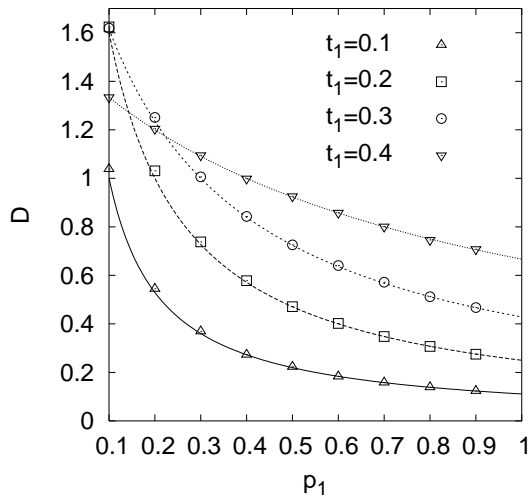}
\caption{
The diffusion constant as a function of $p_1$ and $t_1$
which parametrize $f(t)= p_1 \delta(t-t_1) +(1-p_1) \delta(t-1+t_1) $.
Theoretical and Monte Carlo simulation results are denoted, respectively, by lines and points.}
\label{c1}
\end{figure}

For the binary distribution $f(t)= p_1 \delta(t-t_1) +(1-p_1) \delta(t-1+t_1) $, where $p_1 \in [0.1, 0.2,...,0.9]$ and
$t_1 \in [0.1, 0.2,0.3, 0.4]$, statistics of the photon cloud is evaluated at times
$n \in [7000, 7100,...,10000]$. The mean-square
displacement $\langle  \langle {\bm{j}^2 }\rangle \rangle _{n}$
is computed for each snapshot at
time $n$, and then fitted to $ D n +O$ by the method of linear regression.
An offset $O$ takes
into account the initial ballistic regime.
Fig.\ \ref{c1} shows the excellent agreement between numerical simulations and
$D(p_1,t_1)=t_1(1-t_1)/(p_1-2p_1 t_1 +t_1^2)$, predicted by Eq. (\ref{difema1}).
The maximum differences $ \pm 0.04 $ are comparable to the errorbars $ \pm 0.01 $ which linear regression anticipates.

\begin{figure}
\includegraphics[width=0.85\columnwidth]{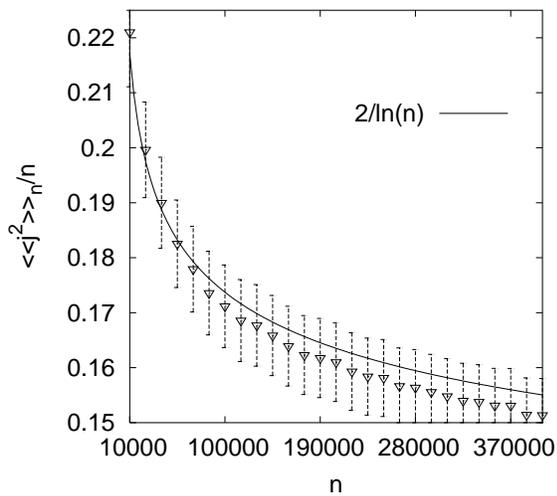}
\caption{$\langle  \langle {\bm{j}^2 }\rangle \rangle _{n}/n$ as function of $n$, for $f(t)=1$ ($0<t<1$).
Theoretical and simulation results are denoted, respectively, by line and points.}
\label{c2}
\end{figure}

\begin{figure}
\includegraphics[width=0.85\columnwidth]{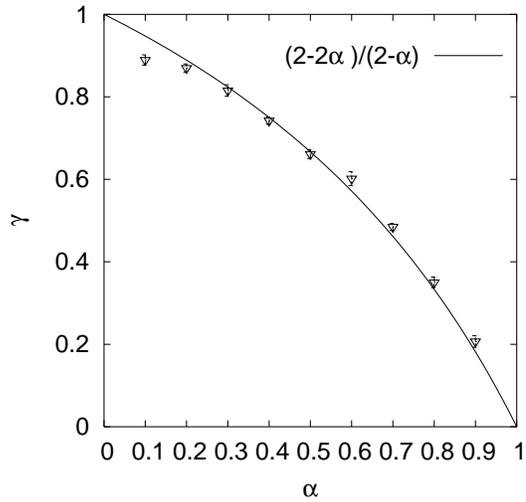}
\caption{$\gamma= \ln[\langle  \langle {\bm{j}^2 }\rangle \rangle _{n}]/\ln[n]$ as $n \rightarrow \infty$, 
versus $\alpha$ which parametrizes $f(t) =(1-\alpha) t^{-\alpha} $.
Theoretical and simulation results are denoted, respectively, by line and points.}
\label{c3}
\end{figure}

For the uniform distribution $f(t)=1$ ($0 \leqslant t \leqslant 1$), statistics is evaluated at times $n \in [10000, 25000, ..., 400000]$.
In Fig.\ \ref{c2}, $\langle  \langle {\bm{j}^2 }\rangle \rangle _{n}/n$, and $2/\ln(n)$ are plotted.
This figure confirms the EMA prediction (\ref{difema2}).

For $f(t) =(1-\alpha) t^{-\alpha} $ ($0 \leqslant t \leqslant 1$), where $\alpha \in [0.1, 0.2,...,0.9]$,
statistics is evaluated at times
$n \in [70000, 71000 ,...,100000]$. $\ln [\langle  \langle {\bm{j}^2 }\rangle \rangle _{n}] $
is computed for each snapshot at time $n$, and then fitted to
$ \gamma \ln(n) +O$, where $O$ is an offset. In Fig.\ \ref{c3}, we plot $\gamma$ and $(2-2\alpha) /(2-\alpha)$ as a
function of $\alpha$, to inspect the EMA prediction (\ref{difema3}). 
The differences are greater than the errorbars which regression anticipates, but the overall agreement
is excellent.

\begin{table*}[t]
\caption{\label{mytable1} EMA prediction of ultimate growth with time of the mean square-displacement.}
\begin{center}
\begin{tabular}{  c c c | c  c c}\hline\hline
 $\text{Photons}$ &  &    & $\text{Excitons}^{\text{a}}$ &  &         \\\hline
& $  f(t)$&     $\langle  \langle {\bm{j}^2 }\rangle \rangle _{n} $ &  & $  f(w)$&     $\langle  \langle {\bm{j}^2 }\rangle \rangle _{\tau} $     \\
$\text{I}$& $<\frac{1}{t}> \neq \infty $&     $n $ &  & $<\frac{1}{w}> \neq \infty  $&     $\tau $     \\
$\text{II}$& $<\frac{1}{t}> = \infty $,   $f(t) \rightarrow f(0)$ $\text{as}$ $t \rightarrow 0 $ &     ${n}/{\ln(n)} $ &  & $<\frac{1}{w}> = \infty  $,   $f(w) \rightarrow f(0)$ $\text{as}$ $w \rightarrow 0 $&     ${\tau}/{\ln{\tau}} $     \\
$\text{III}$& $f(t) \sim  t^{-\alpha} $ $\text{as}$ $t \rightarrow 0 $& $n^{  (2-2\alpha )/(2-\alpha)}$ & &$f(w) \sim  w^{-\alpha} $ $\text{as}$ $w \rightarrow 0 $ & $\tau^{  (2-2\alpha )/(2-\alpha)}$ \\\hline
\end{tabular}
\end{center}
$^{\text{a}}$ From\ \cite{sahimi}. $\tau$ and $w$ denote the continous time and the
transition rate between the nearest neighbours, respectively.
\end{table*}

\section{Discussions and Conclusions}\label{discuss}

In the present paper, we address the persistent random walk on a one-dimensional lattice of {random} transmittances.
The photon transport is diffusive, provided that $<{1}/{t}>$ is finite (class I).
The transmittance of the effective medium is given by Eq. (\ref{tefff}). It expresses the fact
that a few sites with small transmittances (large reflectances) may hinder the photon diffusion.
As percolation properties\ \cite{emabooks}, this feature is induced by the dimensionality of the lattice.
Furthermore, Eq. (\ref{dfg}) shows that fluctuations in transmittances
give rise to a $\sqrt{n}$ correction to the
mean square-displacement. Nevertheless, this correction is negligible at
times $ n \gg \Delta^4/(<1/t>-1)^3 $.

A photon steps back by each reflection.
Intuitively, one expects the abundance of large reflectances to drastically decrease excursion of the photons.
Our EMA predicts a {subdiffusive} transport when $<{1}/{t}>$ is infinite.
The self-consistency equation (\ref{decomposition}) divides such distributions to two classes
II and III, which are summarized in Table\ \ref{mytable1}.
Remarkably, for class III the exponent of mean square-displacement is distribution dependent (nonuniversality).

It would be instructive to compare transport of persistent walkers and
excitation dynamics in random one-dimensional systems\ \cite{sahimi, emabooks, alex}.
Hopping conduction is described by the master equation
\begin{eqnarray}
\frac{\partial P(\tau, \bm{j}) }{\partial \tau} &=& w_{\bm{j}, \bm{j}-1} [P(\tau, \bm{j}-1)- P(\tau, \bm{j})] \nonumber \\
&+ &  w_{\bm{j}, \bm{j}+1}  [P(\tau, \bm{j}+1)- P(\tau, \bm{j})] , \label{exciton}
\end{eqnarray}
where
$0 \leqslant w_{\bm{j}, \bm{j}+1} =w_{\bm{j}+1, \bm{j}}  \leqslant \infty $
is the random transition rate between sites $\bm{j} $ and $\bm{j}+1 $, and $\tau$ is the continous time.
Correlated steps are the essence of (\ref{master1}), while independent steps are the base of (\ref{exciton}).
Despite this celebrated difference, Table\ \ref{mytable1}
reveals remarkable similarities.

As we mentioned in Sec.\ \ref{intro},
the persistent walk on a one-dimensional lattice of random transmittances
arises when the photons move perpendicular to an edge (a face) of the honeycomb (Kelvin) structure.
At normal incidence on a film of random thickness $d_W$ and refractive index $n_W$,
\begin{equation}
t=1-\frac{ 2 r_{W}^2 (1-\cos\beta_W)  }{r_{W}^4-2 r_{W}^2\cos\beta_W+1},
\end{equation}
where $r_{W}=(n_W-1)/(n_W+1)$,
$\beta_W=4 \pi d_W n_W/\lambda $,
and $\lambda$ is the light wavelenght\ \cite{miri2, miri4}. For foams $n_W \sim 1.34$ (water) and $r_{W} \neq \pm 1$, hence
the above transmittance never approaches zero, $<{1}/{t}>$ exists, and the photon transport is diffusive.

In a realization of a distribution $f(t) \sim f_{\alpha} t^{-\alpha} $ as $t \rightarrow 0 $,
small transmittances prevail.
For an experimental observation of the photon subdiffusion, we propose dielectric mirrors
to obtain desired small transmittances. Consider the multilayer cofiguration
$$ \text{H} \text{L} ~\text{H} \text{L}   \cdots \overset{N}{\overbrace{\text{H} \text{L}}}~\text{W}~\text{L}\text{H} \cdots \text{L}\text{H}~\overset{N}{\overbrace{ \text{L}\text{H}}},$$
where $\text{H}$ ($\text{L}$) is a quarter-wave layer with high (low) index $n_H$ ($n_L$),
and $N$ is the number of $\text{H} \text{L} $ ($\text{L}\text{H} $) pairs.
$\text{W}$ is a layer of thickness $d_W$ and index $n_W$.
Following the theory of multilayer films\ \cite{pedro}, we obtain
\begin{equation}
t=1- \frac{({r'}_{W}-1)^2   (1-\cos\beta_W)}{{r'}_{W}^2 (1-\cos\beta_W) +2{r'}_{W}(3+\cos\beta_W )+1-\cos\beta_W  },
\end{equation}
where ${r'}_{W}=n_{W}^2 ({n_H}/{n_L})^{4N}$. 
Apparently, as $N$ increases the transmittance rapidly approaches zero.
For example, suppose $n_H=2.40$ ($\text{TiO}_2$), $n_L=1.38$ ($\text{MgF}_2$), $n_W=1.46$ ($\text{SiO}_2$),
$\lambda \sim 550\text{nm}$ and $\beta_W=\pi$ (quarter-wave layer). For $1$, $2$ and $5$ pairs, the transmittances are
$0.185604$, $0.022174$ and $0.000029$, respectively.
In this realization of small transmittances, two points are considered. First, transmittance of the multilayer is the same
whether light rays go to the right or the left direction. Second, parameters of the multilayer, especially $N$ and
$d_W$, can be varied to fine tune the transmittance. Moreover, absorption of the multilayer is negligible.
Note that thin metalic slabs are highly reflecting, but absorptive.

Our studies can be extended to higher dimensional lattices, and
special two-dimensional photon paths in the honeycomb and Kelvin structures\ \cite{miri1, miri4}.
Another path to pursue is the creation of artificial one-dimensional structures to observe anomalous
diffusion of photons.

\begin{acknowledgments}
We would like to thank M. Sahimi for stimulating discussions and insight to the EMA.
We are grateful to N. Rivier, M. Schmiedeberg, and H. Stark for enlightening comments.
M. F. M and Z. S. appreciate financial support from Iran Telecommunication Research
Center (ITRC).
M. F. M wishes to acknowledge the International Graduate College ``Soft Matter''
at the University of Konstanz, where part of this work is done.
\end{acknowledgments}

\appendix
\section{$z$-transform}\label{zT}
The $z$-transform $F(z)$ of a function $F(n)$ of a discrete variable $n=0, 1, 2,  ...$ is
defined by
\begin{equation}
{F}(z)=\sum_{n=0}^{\infty} F(n) z^n .
\label{ztrans}
\end{equation}
One then derives the $z$-transform of $F(n+1)$ simply as
$F(z)/z - F(n=0)/z$. Note the similarities of this rule with the
Laplace transform of the time derivative of a continuous function\ \cite{weiss, jury}.

Under specified conditions the singular behavior of $F(z)$ can be used to determine the asymptotic behavior of
$F(n)$ for large $n$ (Tauberian theorems)\ \cite{weiss}. For example:
\begin{eqnarray}
 F(z) \sim \frac{\Gamma(1-\alpha)}{(1-z)^{1-\alpha} } &\rightarrow& F(n) \sim \frac{1}{n^{\alpha}} ,\label{Tau1}\\
F(z) \sim\frac{1}{\ln(\frac{1}{1-z})  (1-z)^{\alpha}} &\rightarrow& F(n) \sim \frac{n^{\alpha-1}}{ \Gamma(\alpha) \ln(n) },
\nonumber\\ \label{Tau2}
\end{eqnarray}
where $\Gamma(\alpha)= \int_{0}^{\infty} e^{-t} t^{\alpha-1} dt$.

\section{}\label{apMatrix}
The matrix $\mathbf{M}_4(z,\bm{\theta} )$ introduced in Eq. (\ref{solution}), can be
conveniently written as
\begin{equation}
\mathbf{M}_4(z,\bm{\theta} )  =\begin{pmatrix}  a(z,\bm{\theta} )+b(z,\bm{\theta} ) & a(z,\bm{\theta} )-b(z,\bm{\theta} ) \\
a(z,\bm{\theta} )-b^{*}(z,\bm{\theta} ) & a(z,\bm{\theta} )+b^{*}(z,\bm{\theta} ) \end{pmatrix},
\end{equation}
where
\begin{eqnarray}
 a(z,\bm{\theta} )&=&\frac{1}{2 <\frac{1}{t}>^2}, \nonumber \\
 b(z,\bm{\theta} )&=&\frac{-1}{2 <\frac{1}{t}>}  \sqrt{  \frac{1-z^2}{<\frac{1}{t}>^2- z^2 (<\frac{1}{t}>-2)^2}  }\nonumber \\
&  \times&
\frac{z^2<\frac{1}{t}> -2i z \sin\bm{\theta} - <\frac{1}{t}>  }{z^2(2-<\frac{1}{t}> )  -2z\cos\bm{\theta}  +  <\frac{1}{t}>}.
\end{eqnarray}

\end{document}